# Electron and hole mobility of rutile GeO$_2$ from first principles: an ultrawide-band-gap semiconductor for power electronics

K. Bushick,[1] K. A. Mengle,[1] S. Chae,[1] and E. Kioupakis[1,*]

[1]*Department of Materials Science and Engineering, University of Michigan, Ann Arbor, 48109, United States*

*Correspondence to: kioup@umich.edu

Rutile germanium dioxide (r-GeO$_2$) is a recently predicted ultrawide-band-gap semiconductor with potential applications in high-power electronic devices, for which the carrier mobility is an important material parameter that controls the device efficiency. We apply first-principles calculations based on density functional and density functional perturbation theory to investigate carrier-phonon coupling in r-GeO$_2$ and predict its phonon-limited electron and hole mobilities as a function of temperature and crystallographic orientation. The calculated carrier mobilities at 300 K are $\mu_{\text{elec},\perp \vec{c}} = 244$ cm$^2$ V$^{-1}$ s$^{-1}$, $\mu_{\text{elec},\parallel \vec{c}} = 377$ cm$^2$ V$^{-1}$ s$^{-1}$, $\mu_{\text{hole},\perp \vec{c}} = 27$ cm$^2$ V$^{-1}$ s$^{-1}$, and $\mu_{\text{hole},\parallel \vec{c}} = 29$ cm$^2$ V$^{-1}$ s$^{-1}$. At room temperature, carrier scattering is dominated by the low-frequency polar-optical phonon modes. The predicted Baliga figure of merit of *n*-type r-GeO$_2$ surpasses several incumbent semiconductors such as Si, SiC, GaN, and *β*-Ga$_2$O$_3$, demonstrating its superior performance in high-power electronic devices.

Power electronics are important for the control and conversion of electricity, but inefficiencies cause energy loss during each step in the conversion process, resulting in a combined efficiency of ~80% or less.[1] In the United States, the existing electricity grid is outdated for modern electricity usage and must be replaced with power-conversion electronics that are able to control the power flow more efficiently. Addressing inefficiencies to improve energy sustainability motivates the ongoing search for new materials for power-electronics devices. Ultrawide-band-gap semiconductors with gaps wider than GaN (3.4 eV) have been the focus of power-electronics materials research.[2] Important material parameters to consider for power-electronics applications include the possibility of doping (usually *n*-type, but ambipolar dopability is also desirable for CMOS devices), high carrier mobility $\mu$ for fast switching and low ohmic losses, high thermal conductivity for efficient heat extraction, and a high critical dielectric breakdown field $E_C$ and dielectric constant $\varepsilon_0$ to enable high-voltage operation. The Baliga figure of merit, BFOM = $1/4\, \varepsilon_0 \mu E_C^3$, quantifies the performance of materials in power-electronic devices.[3,4] The BFOM depends most sensitively on the breakdown field, which increases superlinearly with increasing band gap and motivates the search for dopable ultrawide-gap semiconductors.

The $\beta$ polymorph of gallium oxide ($\beta$-Ga$_2$O$_3$) has been the recent focus of attention thanks to the availability of native substrates and the *n*-type dopability with Si or Ge.[2] While its electron mobility is lower than Si, SiC, or GaN, its ultrawide-band-gap of ~4.5 eV produces a high breakdown field and a BFOM superior to these incumbent technologies.[5–9] However, its low thermal conductivity that prevents heat extraction and the unfeasibility of *p*-type doping (due to the formation of self-trapped hole polarons[10]) limit its applicability.[10–12] To overcome these challenges and advance the frontier of power electronics, new ultrawide-band-gap semiconducting materials must be identified and characterized.

Recently, Chae *et al.* found that rutile germanium dioxide (r-GeO$_2$) is a promising ambipolarly dopable semiconductor[13] with an ultrawide band gap (4.68 eV). Donors such as Sb$_{Ge}$ and F$_O$ are shallow (activation energy ~25 meV), while Al$_{Ge}$ and Ga$_{Ge}$ acceptors are deeper with modest ionization energies of ~0.4-0.5 eV.[13] However, the co-incorporation of acceptors with hydrogen and subsequent annealing can produce high acceptor concentrations that exceed the Mott-transition limit and enable *p*-type conduction. Moreover, its measured thermal conductivity of 51 W/m K at 300 K surpasses that of $\beta$-Ga$_2$O$_3$ and facilitates more efficient thermal management.[14] Rutile GeO$_2$ displays similar chemical and structural properties to rutile SnO$_2$, an established *n*-type transparent conductor.[15,16] However, the wider band gap of r-GeO$_2$ is promising for deep-ultraviolet (UV) luminescence and efficient power-electronics applications.[13,17,18] Yet, the electrical conductivity of r-GeO$_2$, and thus its viability and efficiency for power-electronics applications, remain unexplored.

In this work, we apply predictive atomistic calculations to determine the phonon-limited electron and hole mobilities of r-GeO$_2$ as a function of temperature and crystallographic orientation. We quantify the intrinsic phonon and carrier-phonon-coupling properties that impact carrier transport. We find that the room-temperature carrier mobility is dominated by the low-frequency polar-optical phonon modes, and the resulting mobility values at 300 K are $\mu_{\text{elec},\perp\vec{c}}$ = 244 cm$^2$ V$^{-1}$ s$^{-1}$, $\mu_{\text{elec},\|\vec{c}}$ = 377 cm$^2$ V$^{-1}$ s$^{-1}$, $\mu_{\text{hole},\perp\vec{c}}$ = 27 cm$^2$ V$^{-1}$ s$^{-1}$, and $\mu_{\text{hole},\|\vec{c}}$ = 29 cm$^2$ V$^{-1}$ s$^{-1}$. The electron mobility is higher than $\beta$-Ga$_2$O$_3$, while the hole mobility at high temperature surpasses the values for p-type GaN. Our results demonstrate that r-GeO$_2$ exhibits a superior BFOM than current semiconductor technologies such as Si, SiC, GaN, and $\beta$-Ga$_2$O$_3$, and is therefore of great interest for high-power and high-temperature electronics applications.

To accurately predict the carrier and phonon properties of r-GeO$_2$, we use first-principles calculations based on density functional (DFT) and density functional perturbation theories (DFPT) in the local density approximation within Quantum ESPRESSO[19–21] and the iterative Boltzmann transport equation (IBTE) with the EPW code[22–24]. We do not consider spin-orbit coupling. In previous work,[18] we calculated the quasiparticle band structure of r-GeO$_2$ for the experimental lattice parameters[25] using the G$_0$W$_0$ method. We find that the effective masses obtained from G$_0$W$_0$ are similar to those obtained with a hybrid DFT functional (Table SI in the Supplementary Information), verifying the reliability of our calculated band parameters. For phonon calculations, the lattice parameters and atomic positions were relaxed to prevent imaginary phonon frequencies, resulting in lattice parameters of $a$ = 4.516 Å and $c$ = 2.978 Å that differ from experiment[25] by +2.5% and +4.1%, respectively. The phonon dispersion, phonon frequencies at Γ, and sound velocities are included in Fig. S1 and Tables SII and SIII in the Supplementary Information. The quasiparticle energies, phonon frequencies, and carrier-phonon coupling matrix elements were calculated on 4×4×6 Brillouin zone (BZ) sampling grids (using the charge density generated on an 8×8×12 BZ sampling grid for higher accuracy) and interpolated to fine BZ sampling grids with the EPW code. Carrier velocities were evaluated with the velocity operator,[26,27] and the Fröhlich correction was applied to the carrier-phonon coupling matrix elements $g$.[28] The phonon-limited carrier mobilities were calculated over the 100–1000 K temperature range for a carrier concentration of $10^{17}$ cm$^{-3}$.[24] The electron and hole mobilities were converged for carrier and phonon BZ sampling grids of 92×92×138 and 48×48×72, respectively. We sampled states within energy windows of 530 meV around the carrier Fermi energies, which accounts for energy differences during scattering of up to $\hbar\omega_{max} + 5k_BT$ at 1000 K, where $\hbar\omega_{max}$ is the highest optical phonon energy and $k_B$ is the Boltzmann constant. With these calculation parameters, the mobility values at 300 K were converged within 1% for electrons and 2% for holes.

We first discuss the mobility obtained from the iterative solution of the IBTE. The converged carrier mobilities at 300 K are $\mu_{\text{elec},\perp\vec{c}} = 244$ cm$^2$ V$^{-1}$ s$^{-1}$, $\mu_{\text{elec},\|\vec{c}} = 377$ cm$^2$ V$^{-1}$ s$^{-1}$, $\mu_{\text{hole},\perp\vec{c}} = 27$ cm$^2$ V$^{-1}$ s$^{-1}$, and $\mu_{\text{hole},\|\vec{c}} = 29$ cm$^2$ V$^{-1}$ s$^{-1}$. Qualitatively, we expect a lower mobility for transport directions in which carriers have a higher effective mass. Quasiparticle band-structure calculations find that the electron mass is anisotropic ($m^*_{e,\perp\vec{c}} = 0.36\ m_0$ and $m^*_{e,\|\vec{c}} = 0.21\ m_0$) and approximately a factor of 2 lighter for the $\|\ \vec{c}$ direction, which is consistent with the approximately 2 times higher mobility along $\|\ \vec{c}$. We also find that the scattering rates for electrons are relatively isotropic (Fig. S2), indicating that the anisotropy of the calculated electron mobility is primarily driven by the anisotropy of the effective mass. On the other hand, the hole mobility is approximately isotropic, which is consistent with the relatively smaller directional dependence of the effective mass ($m^*_{h,\perp \vec{c}} = 1.29\ m_0$ and $m^*_{h,\|\vec{c}} = 1.58\ m_0$) and scattering rates (Fig. S2).

Figure 1 shows the temperature dependence of the mobility. We can accurately fit the temperature dependence with a combination of exponential equations characteristic of electron-phonon scattering by two optical modes:

$$1/\mu(T) = \frac{1}{\mu_1} e^{\frac{-T_1}{T}} + \frac{1}{\mu_2} e^{\frac{-T_2}{T}} \tag{1}$$

where $\mu_1$, $\mu_2$, $T_1$, and $T_2$ are fitting parameters to describe the temperature dependence. $\mu_1$ and $\mu_2$ are in units of mobility, while $T_1$ and $T_2$ are in units of temperature. Table I lists the fitted values for each carrier type and direction. The two sets of $(\mu_i, T_i)$ correspond to the low- and high-energy polar-optical modes that dominate carrier scattering, as we discuss in greater detail below. Our fitted parameters demonstrate that the low-energy polar-optical modes characterized by the $T_1$ parameter dominate near room temperature. The high-energy polar-optical modes only become important at high temperatures (higher than 500 K) and to a larger extent for electrons than for holes.

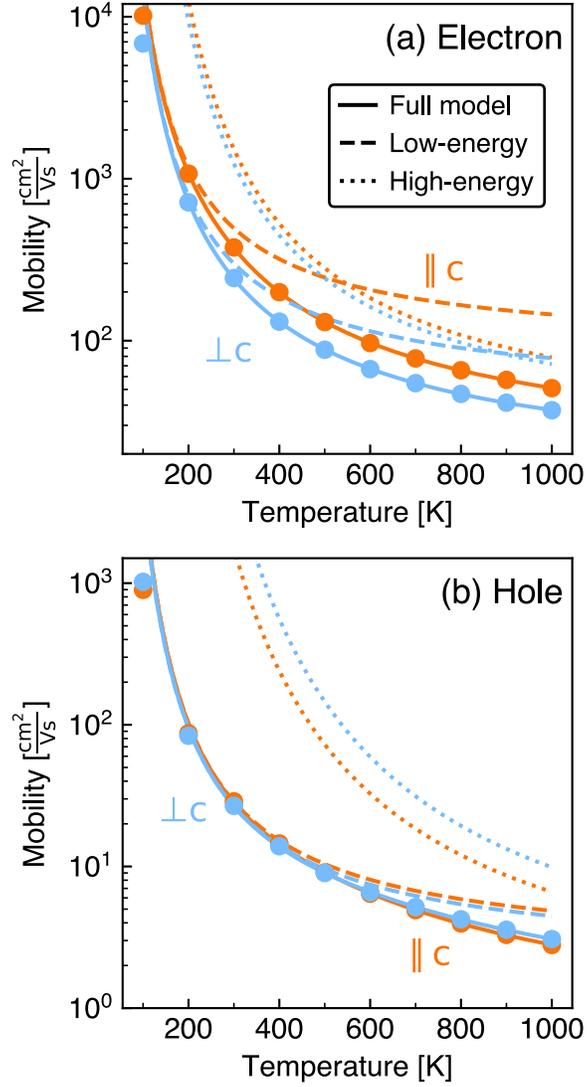

FIG. 1. Electron and hole mobility $\mu$ of r-GeO$_2$ along the $\perp \vec{c}$ and $\parallel \vec{c}$ directions as a function of temperature for a carrier concentration of $n = 10^{17}$ cm$^{-3}$. The solid curves are fits to the data according to Eq. 1 and Table 1. The dashed curves correspond to the low-energy and the dotted curves to the high-energy polar-optical modes.

TABLE I. Fitted parameters for the resistivity model given by $1/\mu(T) = \frac{1}{\mu_1}e^{\frac{-T_1}{T}} + \frac{1}{\mu_2}e^{\frac{-T_2}{T}}$ where $\mu(T), \mu_1,$ and $\mu_2$ are in units of cm² V⁻¹ s⁻¹ and $T, T_1,$ and $T_2$ in K to describe the mobility versus temperature for electrons and holes along the two main crystallographic directions.

| Parameters | Electron, $\perp \vec{c}$ | Electron, $\parallel \vec{c}$ | Hole, $\perp \vec{c}$ | Hole, $\parallel \vec{c}$ |
|---|---|---|---|---|
| $\mu_1$ [cm² V⁻¹ s⁻¹] | 43.8 | 85.9 | 2.08 | 2.27 |
| $T_1$ [K] | 576 | 525 | 765 | 761 |
| $\mu_2$ [cm² V⁻¹ s⁻¹] | 21.1 | 22.2 | 0.669 | 0.603 |
| $T_2$ [K] | 1221 | 1268 | 2691 | 2395 |

Next, we analyze the carrier-phonon coupling matrix elements and scattering lifetimes to understand how different phonon modes affect carrier scattering in r-GeO₂. We first determine the carrier-phonon coupling matrix elements for the bottom conduction and top valence bands for wave vectors along the Γ—X ($\perp \vec{c}$) and Γ—Z ($\parallel \vec{c}$) directions [Fig. 2(a,d)]. Our results show that the polar-optical modes exhibit the strongest carrier-phonon coupling, as expected in polar materials such as r-GeO₂. However, the higher-frequency modes are not as effective at scattering low-energy carriers at room temperature; they either require high temperatures to enable appreciable phonon occupation numbers and scatter carriers by phonon absorption, or high carrier energies to scatter electrons to lower-energy states by phonon emission. Taking the thermal occupation of phonon modes ($n_q$) at room temperature ($k_BT = 26$ meV) into account, we find the dominant modes for carrier scattering by phonon absorption [$g^2 n_q$, Fig. 2(b,e)] and phonon emission [$g^2(n_q + 1)$, Fig. 2(c,f)]. Electrons and holes most strongly absorb the polar-optical modes with frequencies of 41, 49, 94, and 100 meV. As with phonon absorption, carriers scatter most strongly by phonon emission by the polar-optical phonons, as well as the Raman-active modes at 87 meV and 102 meV. We thus predict that carriers primarily scatter by absorption (and to a lesser extent by emission) of the $A_{2u}$ and three $E_u$ polar-optical modes. The eigenmodes of these four dominant scattering phonons are shown in Fig. 2(g-j).

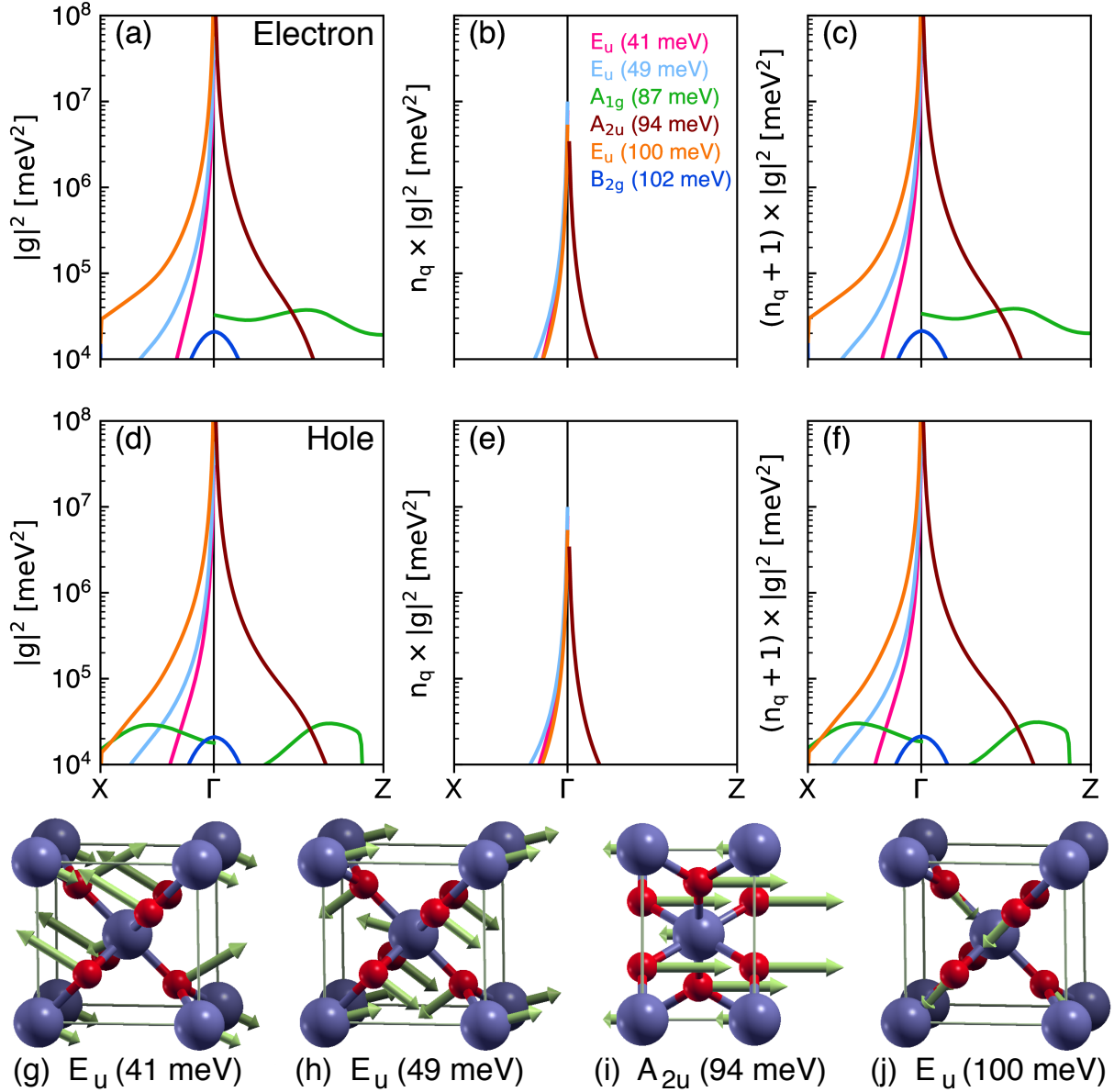

FIG. 2. (a-c) Square of the intraband electron-phonon coupling matrix element $g$ and scattering of electrons via phonon absorption $\left(g^2_{el-ph} n_q\right)$ and phonon emission $\left(g^2_{el-ph}(n_q + 1)\right)$ for the bottom conduction band from $\Gamma$ to $q$ as a function of the phonon wave vector $q$ along the $\Gamma$—X ($\perp \vec{c}$) and $\Gamma$—Z ($\parallel \vec{c}$) directions, showing the phonon modes with the largest coupling strengths.

Phonon occupations are calculated using room temperature ($k_B T = 26$ meV). Panels (d-f) contain the corresponding values for the hole-phonon interactions (i.e., the top valence band). All four IR-active (polar-optical) modes show strong coupling to carriers for wave vectors near $\Gamma$, while the strongest-coupled Raman-active modes show a weak dependence with respect to $q$ (optical deformation potential coupling). (g-j) Atomic displacements corresponding to the dominant polar phonon modes. The larger purple atoms are germanium, and the smaller red atoms are oxygen.

Typically, the polar-optical modes limit the room-temperature mobility in oxide materials, and our electron-phonon coupling results combined with the mobility calculations indicate that this is the case in r-GeO$_2$ as well.[29] Further evidence for the role of the polar-optical modes is provided by the mode-resolved carrier scattering rates as a function of carrier energy (Fig. 3). Here, we refer to modes by their numerical index (in order of increasing frequency) since the directional dependence of the LO-TO splitting reorders the mode frequencies. Mode 7 has the strongest effect at low energies and corresponds to the 41 meV $E_u$ mode. In both the valence band Fig. 3(a) and conduction band Fig. 3(b) there is a noticeable step in the magnitude above 40 meV, where phonon emission becomes possible. The same is true for Mode 17 around 100 meV, which corresponds to both the 94 meV $A_{2u}$ mode and the 100 meV $E_u$ mode, depending on the direction. Modes 9 and 10, which show the next largest contribution to the total electron linewidth are both (depending on the direction) the 49 meV $E_u$ phonon mode. Finally, Mode 13 is associated with the Raman active 67 meV $E_g$ mode, though its contribution is approximately one order of magnitude smaller than the more dominant polar optical modes.

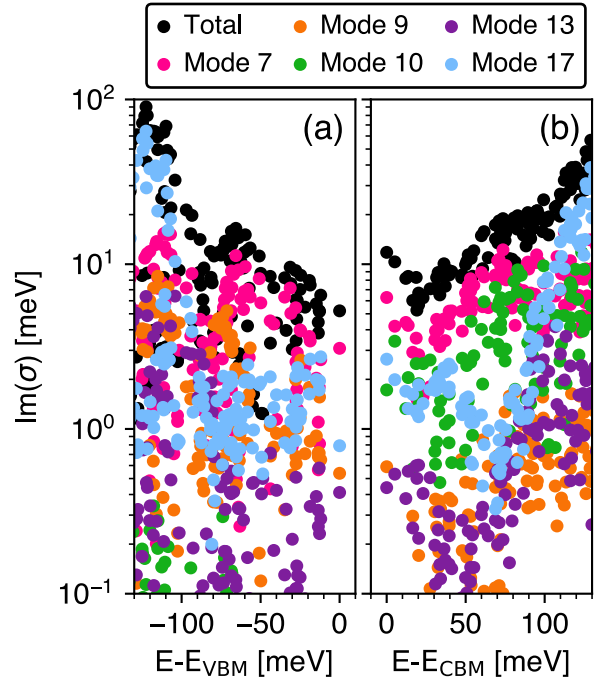

FIG. 3. Phonon-mode-resolved carrier imaginary self-energies (inverse scattering lifetimes) for (a) holes and (b) electrons. The total self-energy (black) is decomposed into the contributions by the dominant phonon modes. Mode 7 corresponds to the 41 meV $E_u$ mode, Modes 9 and 10 correspond to the 49 meV $E_u$ phonon mode, Mode 13 corresponds to the 67 meV $E_g$ mode, and Mode 17 corresponds to the 94 meV $A_{2u}$ mode and the 100 meV $E_u$ mode. The 41 meV $E_u$ mode dominates scattering at low carrier energies (less than approximately 100 meV), while at higher energies scattering by the 94 meV $A_{2u}$ mode and the 100 meV $E_u$ mode dominates.

The carrier mobility is a necessary parameter to understand the performance of a semiconductor in electronic devices. The ultrawide band gap of r-GeO$_2$ (4.68 eV)[17] makes it especially suited for high-power and high-temperature applications.[13,18] Table II lists the material parameters of r-GeO$_2$ relevant for *n*-type and *p*-type power electronics and compares them to incumbent technologies. The breakdown fields of *β*-Ga$_2$O$_3$ (with a gap of 4.5 eV) and r-GeO$_2$ are evaluated using the breakdown field versus band gap relation by Higashiwaki et al.[9] The electron mobility of r-GeO$_2$ is lower than Si, SiC, and GaN by over 70%,[30–32] but higher than $\beta$-Ga$_2$O$_3$. Although the experimental electron mobilities of $\beta$-Ga$_2$O$_3$ have typically been obtained with Hall measurements, it is the drift mobility that should be applied to evaluate the BFOM. A drift mobility of 80 cm$^{-2}$ V$^{-1}$ s$^{-1}$ was measured[33] in $\beta$-Ga$_2$O$_3$ at 300 K, while the highest Hall mobility at 300 K is 184 cm$^{-2}$ V$^{-1}$ s$^{-1}$.[30,34] However, if the Hall factor at 300 K ($r_H = 1.68$)[35] is applied to convert the Hall to drift mobility ($\mu_{\text{Hall}} = \mu_{\text{drift}} r_H$), the highest measured room-temperature drift mobility of $\beta$-Ga$_2$O$_3$ is 109 cm$^{-2}$ V$^{-1}$ s$^{-1}$. Moreover, recent calculations of the phonon-limited carrier mobility of $\beta$-Ga$_2$O$_3$ from first principles using a similar approach as the present work[36] report intrinsic electron and hole drift mobilities at 300 K of 258 cm$^{-2}$ V$^{-1}$ s$^{-1}$ and 1.2 cm$^{-2}$ V$^{-1}$ s$^{-1}$, respectively. Overall, r-GeO$_2$ displays the largest BFOM out of the materials considered here as it exhibits the largest $E_C$ value (since it has the widest band gap), the highest $\varepsilon_0$, and a higher electron mobility than $\beta$-Ga$_2$O$_3$.

Though ultrawide-band-gap materials such as nitrides and $\beta$-Ga$_2$O$_3$ are promising for numerous technological applications, one consistent shortcoming is the relative difficulty of $p$-type doping and low hole mobilities. This is evident in the differences between electron and hole mobility data in Table II. Our findings of a sizable hole mobility in r-GeO$_2$, in combination with the results of Chae *et al.* on the feasibility of ambipolar doping,[13] demonstrate that it may be a candidate material for high-power CMOS devices, in contrast to $\beta$-Ga$_2$O$_3$, which is limited by self-trapped hole polarons.[10] Moreover, the hole mobility of r-GeO$_2$ is comparable to GaN at temperatures below ~500 K, matching or exceeding experimental values and within a factor of two of theoretical values (Fig. S3). At higher temperatures, r-GeO$_2$ shows higher hole mobility than experimentally reported GaN values, despite having a wider band gap. The combination of a higher

BFOM with the prediction of p-type doping and the possibility of efficient hole conduction demonstrate the promise of r-GeO$_2$ as a superior semiconductor compared to incumbent technologies such as β-Ga$_2$O$_3$ and GaN for high-power and high-temperature electronic applications.

TABLE II. Baliga figure of merit for r-GeO$_2$ in comparison to other common ultra-wide-band-gap semiconductors. Electron mobilities and dielectric breakdown fields for all materials are at room temperature and for carrier densities of $10^{16}$ cm$^{-3}$ except those of β-Ga$_2$O$_3$ ($10^{16}$ cm$^{-3}$ and $10^{12}$ cm$^{-3}$ for $\mu_{Hall}$ and $10^{17}$ cm$^{-3}$ for $\mu_{drift}$), GaN ($10^{17}$ cm$^{-3}$ for $\mu_h$), and r-GeO$_2$ ($10^{17}$ cm$^{-3}$).

| Material | Static dielectric constant, $\varepsilon_0$ | Electron mobility, $\mu$ (cm$^2$ V$^{-1}$ s$^{-1}$) | Hole mobility, $\mu$ (cm$^2$ V$^{-1}$ s$^{-1}$) | Dielectric breakdown field, $E_C$ (MV cm$^{-1}$) | n-BFOM ($10^6$ V$^2$ Ω$^{-1}$ cm$^{-2}$) | p-BFOM ($10^6$ V$^2$ Ω$^{-1}$ cm$^{-2}$) |
|---|---|---|---|---|---|---|
| Si | 11.9[37] | 1240 (drift)[31] | | 0.3[9] | 8.8 | |
| 4H-SiC | 9.7[31] | 980 (drift)[31] | | 2.5[9] | 3,300 | |
| GaN | 10.4 (∥ $\vec{c}$)[38] | 1000 (drift)[32] | 31 (Hall)[39] | 3.3[9] | 8,300 | 260 |
| β-Ga$_2$O$_3$ | 10.0[40] | 184; 180 (Hall)[30,34] 109; 80 (drift)[33,34] 258 (drift)[36] | - 1.2 (drift)[36] | 6.4[9] 5.8[36] | 11,000; 10,000 6,300; 4,600 11,000 | 52 |
| r-GeO$_2$ | 14.5 (⊥ $\vec{c}$)[41] | 244 (drift, ⊥ $\vec{c}$) | 27 (drift, ⊥ $\vec{c}$) | 7.0 | 27,000 | 3,000 |
| | 12.2 (∥ $\vec{c}$)[41] | 377 (drift, ∥ $\vec{c}$) | 29 (drift, ∥ $\vec{c}$) | 7.0 | 35,000 | 2,700 |

In summary, we calculate the phonon-limited electron and hole mobilities of r-GeO$_2$ as a function of temperature and crystallographic orientation from first principles, and provide atomistic insights on the dominant carrier-scattering mechanisms. The combination of its ultrawide band gap of 4.68 eV with its predicted electron mobility that is higher than β-Ga$_2$O$_3$ enable a

BFOM that surpasses established power-electronics materials such as Si, SiC, GaN, and $\beta$-Ga$_2$O$_3$. In addition, our results for its sizable hole mobility in combination with the theoretical prediction of its ambipolar dopability[13] indicate potential applications for p-type and CMOS devices. Our results highlight the advantages of r-GeO$_2$ compared to incumbent material technologies for power-electronics applications.

See supplementary material for a comparison of the effective masses calculated with $G_0W_0$ and the hybrid DFT functional, an analytical discussion of the calculated phonon frequencies and sound velocities, the data for the directionally resolved electron self energy due to electron-phonon interaction, and a comparison of the hole mobility of r-GeO$_2$ to available data for GaN.

**AUTHORS' CONTRIBUTIONS**

K.B. and K.A.M. contributed equally to this work.

**Acknowledgements**

This work was supported as part of the Computational Materials Sciences Program funded by the U.S. Department of Energy, Office of Science, Basic Energy Sciences, under Award #DE-SC0020129. It used resources of the National Energy Research Scientific Computing (NERSC) Center, a DOE Office of Science User Facility supported under Contract No. DE-AC02-05CH11231. K.A.M. acknowledges the support from the National Science Foundation Graduate Research Fellowship Program through Grant No. DGE 1256260. K.B. acknowledges the support of the DOE Computational Science Graduate Fellowship Program through grant DE-SC0020347.

**DATA AVAILABILITY**

The data that support the findings of this study are available from the corresponding author upon reasonable request.

**REFERENCES**

[1] K. Afridi, S. Ang, J. Bock, S. Chowdhury, S. Datta, K. Evans, J. Flicker, M. Hollis, N. Johnson, K. Jones, P. Kogge, S. Krishnamoorthy, M. Marinella, T. Monson, S. Narumanchi, P. Ohodnicki, R. Ramesh, M. Schuette, J. Shalf, S. Shahedipour-Sandvik, J. Simmons, V. Taylor, and T. Theis, *Basic Research Needs for Microelectronics* (U.S. Department of Energy, 2018).

[2] J.Y. Tsao, S. Chowdhury, M.A. Hollis, D. Jena, N.M. Johnson, K.A. Jones, R.J. Kaplar, S. Rajan, C.G. Van de Walle, E. Bellotti, C.L. Chua, R. Collazo, M.E. Coltrin, J.A. Cooper, K.R. Evans, S. Graham, T.A. Grotjohn, E.R. Heller, M. Higashiwaki, M.S. Islam, P.W. Juodawlkis, M.A. Khan, A.D. Koehler, J.H. Leach, U.K. Mishra, R.J. Nemanich, R.C.N. Pilawa-Podgurski, J.B. Shealy, Z. Sitar, M.J. Tadjer, A.F. Witulski, M. Wraback, and J.A. Simmons, Adv. Electron. Mater. **4**, 1600501 (2018).

[3] B.J. Baliga, J. Appl. Phys. **53**, 1759 (1982).

[4] B.J. Baliga, IEEE Electron Device Lett. **10**, 455 (1989).

[5] K.A. Mengle, G. Shi, D. Bayerl, and E. Kioupakis, Appl. Phys. Lett. **109**, 212104 (2016).

[6] T. Onuma, S. Saito, K. Sasaki, T. Masui, T. Yamaguchi, T. Honda, and M. Higashiwaki, Jpn. J. Appl. Phys **54**, 112601 (2015).

[7] K.A. Mengle and E. Kioupakis, AIP Adv. **9**, 015313 (2019).

[8] S. Fujita, Jpn. J. Appl. Phys. **54**, 030101 (2015).

[9] M. Higashiwaki, K. Sasaki, A. Kuramata, T. Masui, and S. Yamakoshi, Appl. Phys. Lett. **100**, 013504 (2012).

[10] J.B. Varley, A. Janotti, C. Franchini, and C.G. Van de Walle, Phys. Rev. B **85**, 081109(4) (2012).

[11] M. Higashiwaki and G.H. Jessen, Appl. Phys. Lett. **112**, 060401(4) (2018).

[12] A. Kyrtsos, M. Matsubara, and E. Bellotti, Appl. Phys. Lett. **112**, 032108 (2018).

[13] S. Chae, J. Lee, K.A. Mengle, J.T. Heron, and E. Kioupakis, Appl. Phys. Lett. **114**, 102104(5) (2019).

[14] S. Chae, K.A. Mengle, R. Lu, A. Olvera, N. Sanders, J. Lee, P.F.P. Poudeu, J.T. Heron, and E. Kioupakis, Appl. Phys. Lett. **117**, 102106 (2020).

[15] M. Nagasawa and S. Shionoya, Phys. Lett. **22**, 409 (1966).

[16] A. Schleife, J.B. Varley, F. Fuchs, C. Rödl, F. Bechstedt, P. Rinke, A. Janotti, and C.G. Van de Walle, Phys. Rev. B **83**, 035116(9) (2011).

[17] M. Stapelbroek and B.D. Evans, Solid State Commun. **25**, 959 (1978).

[18] K.A. Mengle, S. Chae, and E. Kioupakis, J. Appl. Phys. **126**, 085703 (2019).

[19] D.M. Ceperley and B.J. Alder, Phys. Rev. Lett. **45**, 566 (1980).

[20] P. Giannozzi, S. Baroni, N. Bonini, M. Calandra, R. Car, C. Cavazzoni, D. Ceresoli, G.L. Chiarotti, M. Cococcioni, I. Dabo, A. Dal Corso, S. De Gironcoli, S. Fabris, G. Fratesi, R. Gebauer, U. Gerstmann, C. Gougoussis, A. Kokalj, M. Lazzeri, L. Martin-Samos, N. Marzari, F. Mauri, R. Mazzarello, S. Paolini, A. Pasquarello, L. Paulatto, C. Sbraccia, S. Scandolo, G. Sclauzero, A.P. Seitsonen, A. Smogunov, P. Umari, and R.M. Wentzcovitch, J. Phys. Condens. Matter **21**, 395502 (2009).

[21] P. Giannozzi, O. Andreussi, T. Brumme, O. Bunau, M. Buongiorno Nardelli, M. Calandra, R. Car, C. Cavazzoni, D. Ceresoli, M. Cococcioni, N. Colonna, I. Carnimeo, A. Dal Corso, S. de Gironcoli, P. Delugas, R.A. DiStasio, A. Ferretti, A. Floris, G. Fratesi, G. Fugallo, R. Gebauer, U. Gerstmann, F. Giustino, T. Gorni, J. Jia, M. Kawamura, H.-Y. Ko, A. Kokalj, E. Küçükbenli, M. Lazzeri, M. Marsili, N. Marzari, F. Mauri, N.L. Nguyen, H.-V. Nguyen, A. Otero-de-la-Roza, L. Paulatto, S. Poncé, D. Rocca, R. Sabatini, B. Santra, M. Schlipf, A.P. Seitsonen, A. Smogunov, I. Timrov, T. Thonhauser, P. Umari, N. Vast, X. Wu, and S. Baroni, J. Phys.

Condens. Matter **29**, 465901 (2017).

[22] S. Poncé, E.R. Margine, C. Verdi, and F. Giustino, Comput. Phys. Commun. **209**, 116 (2016).

[23] F. Giustino, M.L. Cohen, and S.G. Louie, Phys. Rev. B **76**, 165108 (2007).

[24] S. Poncé, E.R. Margine, and F. Giustino, Phys. Rev. B **97**, 121201 (2018).

[25] A.A. Bolzan, C. Fong, B.J. Kennedy, and C.J. Howard, Acta Cryst. **B53**, 373 (1997).

[26] J.R. Yates, X. Wang, D. Vanderbilt, and I. Souza, Phys. Rev. B **75**, 195121 (2007).

[27] X. Wang, J.R. Yates, I. Souza, and D. Vanderbilt, Phys. Rev. B **74**, 195118 (2006).

[28] C. Verdi and F. Giustino, Phys. Rev. Lett. **115**, 176401 (2015).

[29] A. Samanta, M. Jain, and A.K. Singh, J. Chem. Phys. **143**, 064703(6) (2015).

[30] Y. Zhang, A. Neal, Z. Xia, C. Joishi, J.M. Johnson, Y. Zheng, S. Bajaj, M. Brenner, D. Dorsey, K. Chabak, G. Jessen, J. Hwang, S. Mou, J.P. Heremans, and S. Rajan, Appl. Phys. Lett. **112**, 173502 (2018).

[31] B.J. Baliga, *Fundamentals of Power Semiconductor Devices* (Springer Science & Business Media, Berlin, 2010).

[32] M.E. Levinshtein, S.L. Rumyantsev, and M.S. Shur, *Properties of Advanced Semiconductor Materials: GaN, AlN, InN, BN, SiC, SiGe* (John Wiley & Sons, Hoboken, NJ, 2001).

[33] A. Parisini and R. Fornari, Semicond. Sci. Technol. **31**, 035023 (2016).

[34] Z. Feng, A.F.M. Anhar Uddin Bhuiyan, M.R. Karim, and H. Zhao, Appl. Phys. Lett. **114**, (2019).

[35] N. Ma, N. Tanen, A. Verma, Z. Guo, T. Luo, H. (Grace) Xing, and D. Jena, Appl. Phys. Lett. **109**, 212101(5) (2016).

[36] S. Poncé and F. Giustino, Phys. Rev. Res. **2**, 033102 (2020).

[37] S. Sze and K.N. Kwok, in *Phys. Semicond. Devices*, 3rd ed. (John Wiley & Sons, Hoboken,

NJ, 2007), p. 549.

[38] J.A.S. Barker and M. Ilegems, Phys. Rev. B **7**, 743 (1973).

[39] M. Horita, S. Takashima, R. Tanaka, H. Matsuyama, K. Ueno, M. Edo, T. Takahashi, M. Shimizu, and J. Suda, Jpn. J. Appl. Phys. **56**, 031001 (2017).

[40] K. Sasaki, A. Kuramata, T. Masui, E.G. Víllora, K. Shimamura, and S. Yamakoshi, Appl. Phys. Express **5**, 035502 (2012).

[41] D.M. Roessler and W.A. Albers, J. Phys. Chem. Solids **33**, 293 (1972).

# Supplementary Material for:

## "Electron and hole mobility of rutile GeO$_2$ from first principles: an ultrawide-band-gap semiconductor for power electronics"

Kyle Bushick,[1] Kelsey A. Mengle,[1] Sieun Chae, and Emmanouil Kioupakis*

Department of Materials Science and Engineering, University of Michigan, Ann Arbor, Michigan 48109, USA

*E-mail: kioup@umich.edu

**I. Comparison between G$_0$W$_0$ and HSE effective masses**

The carrier mobility is intrinsically linked to the carrier effective masses, which is in turn dependent on the calculated band structure. In order to further establish confidence in the reliability in our results, we show that the differences between the effective masses calculated from the HSE band structure and the G$_0$W$_0$ band structure are relatively minor. Effective masses are calculated via fitting a hyperbolic equation according to Mengle et al.,[1] and are shown in Table S1. The HSE band structure was calculated as in Chae et al..[2]

TABLE SI. Comparison of directionally dependent electron and hole effective masses fit to the band structure calculated via the G$_0$W$_0$ and HSE methods.

|  | Γ—X (electron) | Γ—Z (electron) | Γ—X (hole) | Γ—Z (hole) |
|---|---|---|---|---|
| G$_0$W$_0$ | 0.36 | 0.21 | 1.29 | 1.58 |
| HSE | 0.31 | 0.23 | 1.09 | 1.57 |

**II. Phonon properties**

The phonon dispersion of a material strongly affects the carrier mobilities. Figure S1 shows the wave vector dependence of each mode along the Γ—X ($\perp \vec{c}$) and Γ—Z ($\parallel \vec{c}$) directions. The lowest-frequency optical mode (B$_{1g}$, 98 cm$^{-1}$) occurs at Γ. Along the in-plane direction, all phonon modes with frequencies above 179 cm$^{-1}$ are optical modes, while the acoustic modes extend to a higher frequency (391 cm$^{-1}$) in the out-of-plane direction. Table SII lists the phonon frequencies of all modes, including the transverse optical (TO) and longitudinal optical (LO) splitting for each of the four infrared (IR)-active modes. Also shown is additional data from two computational studies and two experimental studies. Our calculated phonon frequencies are in overall good agreement with previous experimental and theoretical reports. An exception is the TO/LO modes reported by Kahan *et at*. at 652/680 cm$^{-1}$,[3] which have not been reproduced in other experimental or theoretical reports.[4,5] A notable discrepancy also occurs for the lowest-frequency (98 cm$^{-1}$) of the Raman-active $B_{1g}$ mode at Γ, which Kaindl *et al.* calculated at 182 cm$^{-1}$ for a unit-cell volume of 56.5 Å$^3$.[5] However, Samanta *et al.* show that of both the band gap and the phonon frequencies of r-GeO$_2$ are very sensitive to the unit cell volume and calculated a $B_{1g}$ frequency (109 cm$^{-1}$) similar to ours.[6] This sensitivity explains the difference between our calculated $B_{1g}$ mode frequency and experiment.[7] Table SIII lists the values of the sound velocity of r-GeO$_2$ in the $\perp \vec{c}$ and $\parallel \vec{c}$ directions for each acoustic mode and are compared to experimental measurements of the sound velocities derived from the elastic constants.[8]

---

[1]these authors contributed equally

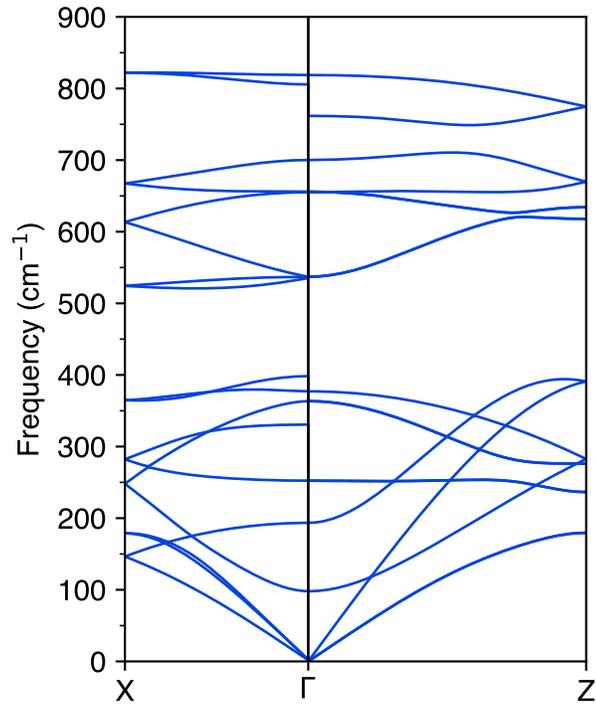

FIG. S1. Phonon dispersion of r-GeO$_2$ along the $\Gamma$—X ($\perp \vec{c}$) and $\Gamma$—Z ($\parallel \vec{c}$) directions, including LO-TO splitting.

TABLE SII. Calculated phonon frequencies (in cm$^{-1}$) at $\Gamma$ for r-GeO$_2$. The activity of each mode is indicated as R (Raman-active), IR (infrared-active), or — (silent). The TO and LO frequencies

for the IR-active modes, including the direction of splitting, are indicated. Our theoretical values are compared to other computed results by (a) Samanta et al.[6] and (b) Kaindl et al.[5] as well as experimental results by (c) Kaindl et al.[5], and (d) Kahan et al.[3]

| Mode Type | Activity | Present work | (a)[6] | (b)[5] | (c)[5] | (d)[3] |
|---|---|---|---|---|---|---|
| | | Calculated | | | Experimental | |
| $B_{1g}$ | R | 98 | 109 | 182 | 170 | |
| $B_{1u}$ | — | 194 | 211 | 219 | | |
| $E_u$ | I (TO) | 252 | 293 | 317 | | 334 |
| $E_u$ | I (LO, $\Gamma$—X) | 331 | | 362 | | 484 |
| $E_u$ | I (TO) | 363 | 383 | 364 | | 652 |
| $E_u$ | I (LO, $\Gamma$—X) | 398 | | 474 | | 680 |
| $A_{2g}$ | — | 377 | 447 | 479 | 476 | |
| $A_{2u}$ | I (TO) | 535 | 543 | 521 | 496 | 522 |
| $A_{2u}$ | I (LO, $\Gamma$—Z) | 762 | | 801 | 792 | 816 |
| $E_g$ | R | 537 | 549 | 546 | | |
| $B_{1u}$ | — | 655 | 553 | 672 | 680, 687 | |
| $E_u$ | I (TO) | 656 | 679 | 665 | 648 | 709 |
| $E_u$ | I (LO, $\Gamma$—X) | 806 | | 837 | 849 | 852 |
| $A_{1g}$ | R | 700 | 693 | 711 | 700 | |
| $B_{2g}$ | R | 819 | 880 | 869 | 874 | |

TABLE SIII. Calculated sound velocities (in km/s) of r-GeO$_2$ along the $\Gamma$—X ($\perp \vec{c}$) and $\Gamma$—Z ($\parallel \vec{c}$) directions for each acoustic phonon branch. We show experimental sound velocities derived from the elastic constant measurements by Wang and Simmons for comparison.[8]

| Direction | (Bottom) TA | (Top) TA | LA | Source |
|---|---|---|---|---|
| $\Gamma$—X | 4.70 | 6.60 | 6.74 | This work |
| $\Gamma$—X | | 6.415 | 7.328 | Ref. [8] |
| $\Gamma$—Z | 4.67 | 4.67 | 9.44 | This work |
| $\Gamma$—Z | | 5.072 | 9.770 | Ref. [8] |

### III. Directionally resolved self-energy

We further select only wave vectors that lie along the high-symmetry Brillouin-zone paths to examine the directional anisotropy of the scattering rate (Fig. S2). Within 100 meV of the band edges, there is no clear anisotropy for either the hole or the electron scattering rates. However, at higher energies the holes show some anisotropy, with points along the Γ—Z direction showing larger values than those in the Γ—X direction (symmetrically equivalent to Γ—Y). The electrons, on the other hand, show no such anisotropy. We note that the anisotropy occurs at carrier energies far from the band edges, which do not impact the carrier mobilities.

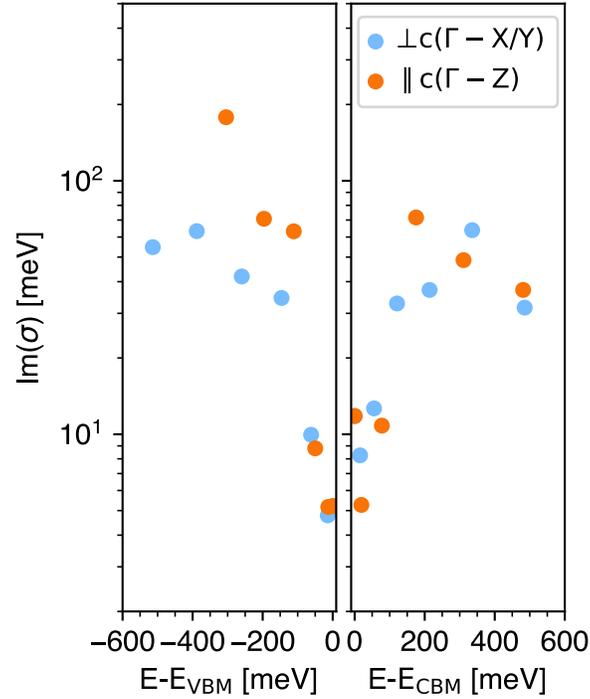

FIG. S2. Directionally resolved carrier imaginary self-energies due to electron-phonon interaction. These points are a subset of the points present in Fig. 3 that have coordinates along the Γ—X, Γ—Y, and Γ—Z directions.

**IV. Comparison of hole mobility to GaN**

We compare the hole mobility of r-GeO$_2$ to three previous reports for GaN in Fig. S3. The data from Horita *et al.* are from Hall measurements performed on a GaN sample with an Mg concentration of $6.5\times10^{16}$ cm$^{-3}$ leading to $N_A = 7.0\times10^{16}$ cm$^{-3}$ and $N_D = 3.2\times10^{16}$ cm$^{-3}$.[9] The data from Kozodoy *et al.* are from Hall measurements performed on a GaN sample with an Mg concentration of $1.6\times10^{19}$ cm$^{-3}$ leading to $N_A = 1.8\times10^{19}$ cm$^{-3}$ and $N_D = 1.1\times10^{18}$ cm$^{-3}$.[10] The data from Poncé *et al.* are for the phonon-limited mobility in intrinsic GaN and is calculated with a similar methodology to our work.[11]

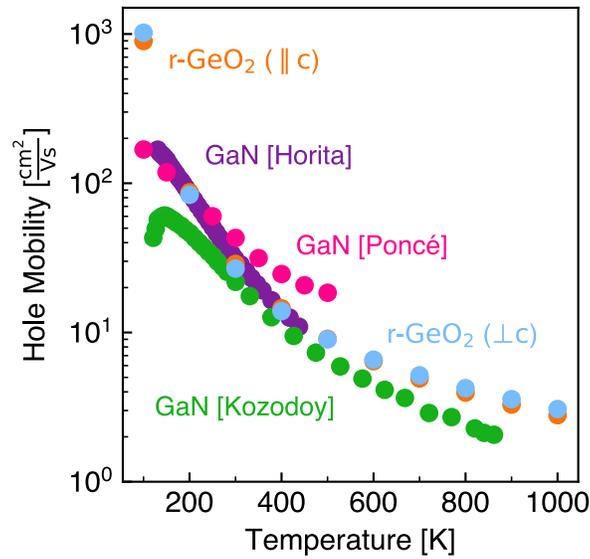

FIG. S3. Our calculated hole mobility of r-GeO$_2$ compared to two experimental reports of the hole mobility in *p*-type GaN with $N_A = 7.0\times10^{16}$ cm$^{-3}$ for the Horita et al. data[9] and $N_A = 1.8\times10^{19}$ cm$^{-3}$ for the Kozodoy et al. data,[10] and to one theoretical report of the phonon-limited hole mobility by Poncé et al.[11]

**References**

[1] K.A. Mengle, S. Chae, and E. Kioupakis, J. Appl. Phys. **126**, 085703 (2019).

[2] S. Chae, J. Lee, K.A. Mengle, J.T. Heron, and E. Kioupakis, Appl. Phys. Lett. **114**, 102104(5)

(2019).

[3] A. Kahan, J.W. Goodrum, R.S. Singh, and S.S. Mitra, J. Appl. Phys. **42**, 4444 (1971).

[4] M. Madon, P. Gillet, C. Julien, and G.D. Price, Phys. Chem. Miner. **18**, 7 (1991).

[5] R. Kaindl, D.M. Többens, S. Penner, T. Bielz, S. Soisuwan, and B. Klötzer, Phys. Chem. Miner. **39**, 47 (2012).

[6] A. Samanta, M. Jain, and A.K. Singh, J. Chem. Phys. **143**, 064703(6) (2015).

[7] A.A. Bolzan, C. Fong, B.J. Kennedy, and C.J. Howard, Acta Cryst. **B53**, 373 (1997).

[8] H. Wang and G. Simmons, J. Geophys. Res. **78**, 1262 (1973).

[9] M. Horita, S. Takashima, R. Tanaka, H. Matsuyama, K. Ueno, M. Edo, T. Takahashi, M. Shimizu, and J. Suda, Jpn. J. Appl. Phys. **56**, 031001 (2017).

[10] P. Kozodoy, H. Xing, S.P. DenBaars, U.K. Mishra, A. Saxler, R. Perrin, S. Elhamri, and W.C. Mitchel, J. Appl. Phys. **87**, 1832 (2000).

[11] S. Poncé, D. Jena, and F. Giustino, Phys. Rev. Lett. **123**, 096602 (2019).